\documentclass[preprint]{elsarticle}

\usepackage{amsmath}
\usepackage{lineno,hyperref}
\usepackage[version=3]{mhchem}
\usepackage{graphicx}
\usepackage{epstopdf}
\graphicspath{{Figures/}}
\modulolinenumbers[5]
\usepackage{xcolor}
\usepackage{siunitx}
\usepackage{textcomp}

\journal{Journal of \LaTeX\ Templates}

\newcommand\hl[1]{%
  \bgroup
  \hskip0pt\color{red!80!black}%
  #1%
  \egroup
}

\newif\ifhighlight
\highlighttrue 
\ifhighlight
    
\else
    \renewcommand{\hl}[1]{#1}
\fi

\usepackage{subfig}

\begin{document}

\begin{frontmatter}

\title{Smart cathodic protection system for real-time quantitative assessment of corrosion of sacrificial anode based on Electro-Mechanical Impedance (EMI)}
\tnotetext[mytitlenote]{D. Tamhane and J. Thalapil have contributed equally to the manuscript.}

\author[1]{Durgesh Tamhane}
\author[1]{Jeslin Thalapil}

\author[2]{Sauvik Banerjee}
\ead{sauvik@civil.iitb.ac.in}

\author[1]{Siddharth Tallur\corref{cor}}
\ead{stallur@iitb.ac.in}

\address[1]{Department of Electrical Engineering, Indian Institute of Technology Bombay, Mumbai 400076, MH India}
\address[2]{Department of Civil Engineering, Indian Institute of Technology Bombay, Mumbai 400076, MH India}
\cortext[cor]{Corresponding author}

\begin{abstract}

Corrosion of metal structures is often prevented using cathodic protection systems, that employ sacrificial anodes that corrode more preferentially relative to the metal to be protected. In-situ monitoring of these sacrificial anodes during early stages of their useful life could offer several insights into deterioration of the material surrounding the infrastructure as well as serve as early warning indicator for preventive maintenance of critical infrastructure.
In this paper, we present an Electro-Mechanical Impedance (EMI) measurement-based technique to quantify extent of corrosion of a zinc sacrificial anode without manual intervention. 
The detection apparatus consists of a lead zirconate titanate (PZT) transducer affixed onto a circular zinc disc, with waterproofing epoxy protecting the transducer element when the assembly is submerged in liquid electrolyte (salt solution) for accelerated corrosion by means of impressed current. We develop an analytical model for discerning the extent of corrosion by monitoring shift in resonance frequency for in-plane radial expansion mode of the disc, that also accurately models the nonlinearity introduced by partial delamination of the corrosion product (zinc oxide) from the disc. 
The analytical model thus developed shows excellent agreement with Finite Element Analysis (FEA) and experimental results.
Our work establishes the efficacy of the proposed technique for monitoring the state of health of sacrificial anodes in their early stage of deterioration and could thus be widely adopted for structural health monitoring applications within the internet of things.
\end{abstract}

\begin{keyword}
Electro-Mechanical Impedance (EMI) \sep sacrificial anode \sep cathodic protection system \sep corrosion monitoring \sep smart infrastructure \sep structural health monitoring
\end{keyword}

\end{frontmatter}


\section{Introduction}
\label{intro}
Longevity of civil infrastructure relies on strength of metals such as steel for construction of buildings, pipelines, storage tanks etc. These metals are typically embedded in insulated casings or treated with anti-corrosion coatings to protect them from damage due to corrosion.
Nevertheless, these metals are susceptible to corrosion as soon as the material in which they are embedded deteriorates and allows permeation and diffusion of corrosive agents. The initiation and rate of corrosion depends on several factors e.g. quality of 
raw materials used in preparation of concrete for reinforced steel structures, air quality of the environment where the structure is to be built, water quality and salinity for marine infrastructure etc. 
A simple and commonly used method to protect metals from corrosion is by installing a cathodic protection system 
\cite{bertolini2002prevention,szabo2006cathodic,cicek2013cathodic}. 
The cathodic protection system uses a sacrificial metal that is more electrochemically active (i.e. more negative electrode potential) relative to the vulnerable metal that requires protection. In passive cathodic protection systems, the sacrificial metal (anode) is attached to the vulnerable metal surface prone to corrosion. Being more electrochemically active, the sacrificial anode is consumed due to corrosion in place of the metal that it protects, thus prolonging the service life of the infrastructure.
It is necessary to monitor the efficacy of a passive cathodic protection system to ensure adequate protection against corrosion.
This is often measured indirectly 
by monitoring the corrosion of the metal structure that the cathodic system is protecting \cite{brousseau1998laboratory,jeong2013electrochemical}.  
Other direct monitoring methods include half-cell potential measurement \cite{half_cell_astuti2019effectiveness}, open-circuit potential test \cite{open_potential_angst2019critical}, DC current density measurement \cite{DC_current_density_farooq2019evaluating} etc.
These tests require three-electrode apparatus and disconnection of the sacrificial anodes from the cathodic protection system for several hours thus interfering with their functionality.

The advent of the internet of things has lead to tremendous surge in smart infrastructure applications within structural health monitoring.
Lead zirconate Titanate (PZT) transducers find a variety of applications in structural health monitoring such as 
assessing performance and state of pre-stressed concrete \cite{ai2019numerical,kaur168cost}, damage localization and monitoring \cite{ostachowicz2009damage, huynh2019sensing,zhu146electromechanical,mandal2019identification}, monitoring debonding in composite structures \cite{chen2019debonding,xu2018dominance,aslam2020dynamic}, monitoring of fatigue damage and curing of adhesives \cite{tang2019development, haq2020fatigue} etc. More recently, PZT transducers embedded within infrastructure are employed as smart sensors for real-time monitoring \cite{maurya2020smart,jiao2020piezoelectric} for applications such as impact monitoring \cite{qiu2020mechatronic} and detection of internal damage in concrete and embedded reinforcements \cite{sriramadasu2019identification,kocherla2020embedded}.
While PZT transducers have also been employed for corrosion monitoring \cite{wang2020electromechanical, shi2019corrosion}, an analytical and quantitative correlation of change in impedance of the PZT transducer due to the corrosion of the system under inspection requires further development. More recently, Electro-Mechanical Impedance (EMI) measurements of PZT transducers have been explored for monitoring the performance of concrete including monitoring of chlorine ingress and carbonation \cite{talakokula2018monitoring,zhang2020monitoring,talakokula2015reinforcement,talakokula2016diagnosis} of concrete. EMI measurements conducted with PZT transducers embedded into stainless steel coupons \cite{li2019pzt} have also been reported for studying the effect of corrosion on their conductance spectra. While the versatility of EMI measurements with PZT transducers is well established in literature, their application as a tool for health-assessment of cathodic protection systems has not been investigated.

\begin{figure}[!tbp]
    \centering
    \includegraphics[width =\linewidth ]{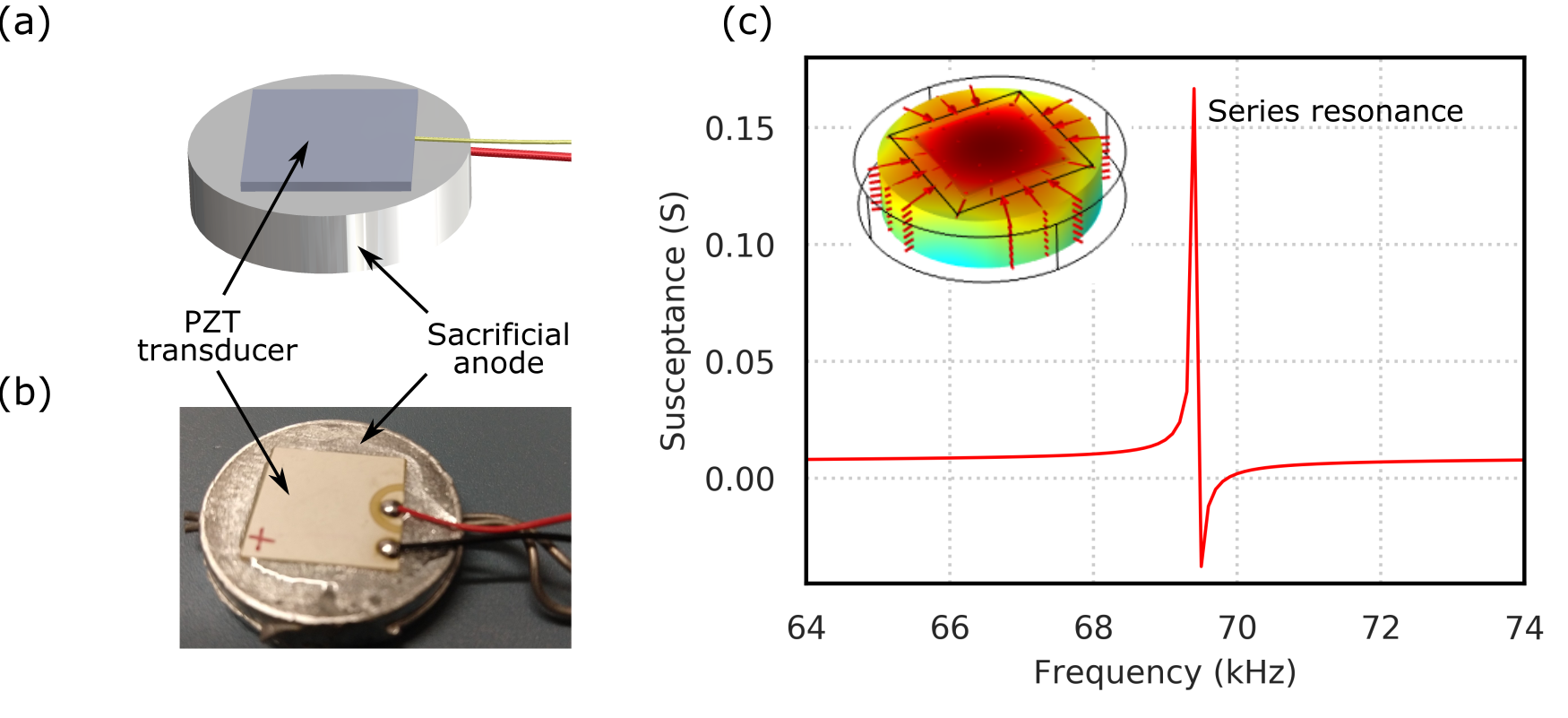}
    \caption{(a) 3D illustration, and (b) photograph of the PZT transducer bonded to zinc sacrificial anode. (c) Finite element method simulation of susceptance spectrum of the sensor showing electro-mechanical resonance corresponding to radial expansion mode of the assembly (inset: mode-shape showing displacement profile for radial expansion mode). The extent of corrosion of the sacrificial anode is ascertained by monitoring the shift in series resonance frequency.}
    \label{fig:structure_mode}
\end{figure}

In this paper, we report a novel EMI-based sensor topology for monitoring the corrosion of a sacrificial zinc anode with an adhesively bonded PZT transducer. Experimental measurements are conducted with an impressed-current based accelerated corrosion system for in-situ evaluation of the susceptance spectra of the PZT transducer attached to the sacrificial anode. Corrosion-induced change in frequency for a bulk acoustic resonance mode of the disc-shaped sacrificial anode is employed as a signature to assess the extent of corrosion as shown in Figure~\ref{fig:structure_mode}. 
We observe that the series resonance frequency of the PZT transducer gradually increases with increasing corrosion time. Detailed Finite Element (FE) simulations and analytical models are presented in an effort to examine this shift in frequency, which is attributed to the formation and partial delamination of the oxide film formed on the surface of the sacrificial anode.
The following sections present details of the analytical model (with supporting information in appendices), and experimental results and discussions.

\section{Analytical Model and Finite Element Analysis}
\subsection{Principle of operation}

The coupling of the mechanical parameters (mechanical strain: $S_{ij}$, mechanical stress: $T_{kl}$) and electrical parameters (electrical field: $E_{kl}$, electrical displacement: $D_{j}$) for a structure excited with piezoelectric transduction can be described as \cite{VictorGiurgiutiu}:
 
\begin{equation}
\begin{split}
S_{i j}=s_{i j k l}^{E} T_{k l}+d_{k i j} E_{k} \\
D_{j}=d_{j k l} T_{k l}+\varepsilon_{j k}^{T} E_{k}
\end{split}
\end{equation}
where $s_{ijkl}^{E}$ is the mechanical compliance of the structure material measured at zero electric field $(E=0)$, $\varepsilon_{jk}^{T}$ is the dielectric permittivity measured at zero mechanical stress $(T=0)$, and $d_{kij}$ represents the piezoelectric coupling coefficient. The direct piezoelectric effect describes conversion of applied stress to electric
charge. Similarly, the converse piezoelectric effect results in generation of strain under application of an electric field in the piezoelectric material. At electric resonance, the PZT transducer presents low impedance (i.e. large admittance) to the excitation source, which results in large mechanical response when coupled to a mechanical resonance of the structure.
The electro-mechanical resonance frequency and impedance of the piezoelectrically transduced structure depends on its geometry, mass and stiffness. Corrosion causes change in these parameters, thereby resulting in shift in the electro-mechanical resonance frequency as seen in the EMI spectrum.
Using an impedance analyzer/LCR meter, the resonance frequency of the structure can be measured and used to determine the extent of corrosion. Here we consider a circular zinc plate as the sacrificial anode, onto which a PZT transducer is bonded using an adhesive. The resonance mode under study is the radial expansion mode as shown in Figure~\ref{fig:structure_mode}(c). Corrosion of zinc (Zn) results in formation of a zinc oxide layer (ZnO) on the anode, which results in change in the series resonance frequency, corresponding to the peak in the admittance spectrum. We present below an analytical model for the radial expansion mode of vibration for the multi-layer (Zn, ZnO) disc.

\subsection{Mathematical formulation for axial in-plane vibration of circular plates}

The governing equation for in-plane axisymmetric radial expansion mode of a circular plate in cylindrical coordinates ($r$,$\theta$) is expressed as \cite{VictorGiurgiutiu}:

\begin{equation}
    \frac{E}{1-\nu^2}\left(\frac{\partial^2 u_r}{\partial r^2}+\frac{1}{r}\frac{\partial u_r}{\partial r}-\frac{u_r}{r^2}\right)=\rho\frac{\partial^2 u_r}{\partial t^2}
\end{equation}
where $E$ and $\nu$ are the Young's modulus and Poisson's ratio of the material respectively, $\rho$ is the density and $u_r$ is the radial displacement in cylindrical coordinates ($r$,$\theta$).
The general solution for the radial displacement $u_r$ is expressed as (details provided in \ref{app:a}):

\begin{equation}
    u_r\left(r,t\right)=\left(C_1 J_1\left(\beta\frac{r}{a}\right)+D_1 Y_1\left(\beta\frac{r}{a}\right)\right)e^{i\omega t}
\end{equation}
where $\omega$ denotes the fundamental resonance frequency of the circular plate of radius $a$ and $\beta=\omega\sqrt{\frac{\rho\left(1-\nu^2\right)a^2}{E}}$ denotes dimensionless frequency parameter, $J_1(\beta\frac{r}{a})$ is Bessel function of first kind of order one, and $Y_1(\beta\frac{r}{a})$ is Bessel function of second kind of order one.
The above solution does not consider the effect of inertia of lateral motion. Thus, similar to Rayleigh's theory for the extensional vibration of bars, the modified frequency parameter for a circular disc can be expressed as:
\begin{equation}
    \beta=\omega\sqrt{\frac{\rho a^2}{\frac{E}{1-\nu^2}-\frac{\rho \omega^2 \nu^2 I_{zz}}{A}}}
\end{equation}
where $I_{zz}$ denotes the area moment of inertia of the disc cross-section along a direction normal to the circular cross-section, and $A$ denotes area of the cross-section. The detailed formulation for dimensionless frequency parameter for a thick circular disc is provided in \ref{app:b}.

\subsection{Equivalent circular-shaped PZT model}

We have employed a square-shaped PZT transducer in our experiments, as shown in Figure~\ref{fig:structure_mode}. To simplify the mathematical formulation, we model the square PZT transducer as an equivalent circular-shaped PZT transducer so that interface boundary condition in cylindrical coordinates can be established.
For a square plate undergoing axial vibration with `free edge' boundary condition, the fundamental frequency is expressed as~\cite{VictorGiurgiutiu}:
\begin{equation}\label{eqn:sq_T}
    \omega=\frac{\pi}{L}\sqrt{\frac{E_T}{1-\nu_T^2}}
\end{equation}
where $L$ denotes the edge length of the square PZT transducer, and $E_T$ and $\nu_T$ denote the Young's modulus and Poisson's ratio of the transducer material (PZT).
Similarly, for a circular plate undergoing axial vibration with similar boundary conditions, the fundamental frequency is expressed as:
\begin{equation}\label{eqn:circ_T}
    \omega=\frac{\beta_T}{a_T}\sqrt{\frac{E_T}{1-\nu_T^2}}
\end{equation}
where $a_T$ is the radius of circular PZT transducer and $\beta_T$ is the dimensionless frequency parameter obtained by solving the characteristic equation 
\begin{equation}\label{eq7}
    \beta_T  J_0(\beta_T )-(1- \nu_T) J_1(\beta_T )=0 
\end{equation}
where $J_0$ and $J_1$ are the Bessel functions of first kind of order $0$ and $1$, respectively \cite{VictorGiurgiutiu}.

Equating equations~\eqref{eqn:sq_T} and \eqref{eqn:circ_T}, we obtain
\begin{equation}
    \frac{\pi}{L}=\frac{\beta_T}{a_T}
\end{equation}
or,
\begin{equation}\label{eqn:eqv_radius}
    a_T=L\frac{\beta_T}{\pi}
\end{equation}

\begin{figure}[!htbp]
    \centering
    \includegraphics[width = 1\linewidth]{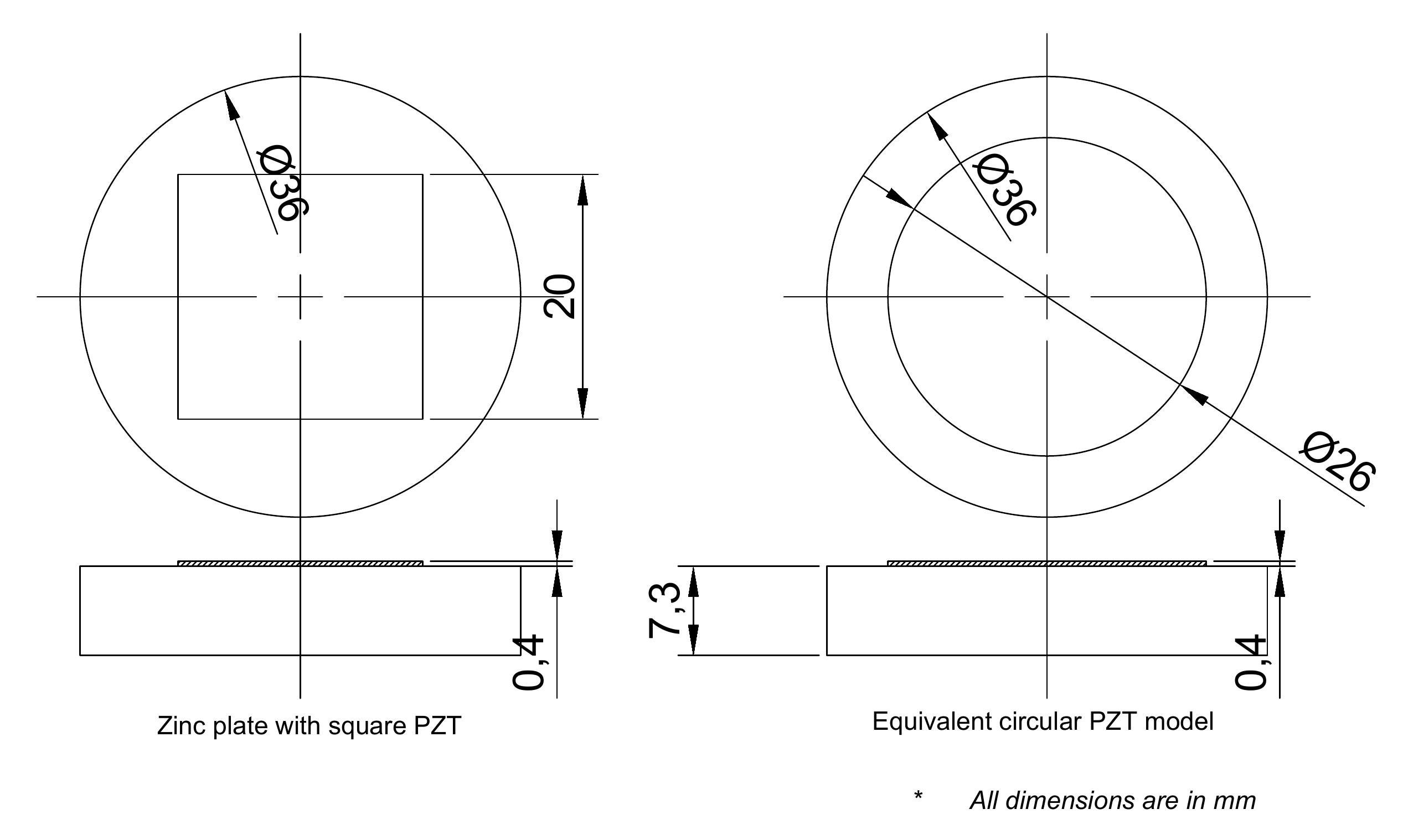}
    \caption{Dimensions of (left) square-shaped PZT transducer atop zinc anode used in our experiment, and (right) equivalent circular-shaped PZT transducer with dimensions computed using equation~\eqref{eqn:eqv_radius}. All dimensions are in units of \SI{}{\mm}.}
    \label{fig:eqv_circ}
\end{figure}

The value of $\beta_T$ obtained by solving the characteristic equation (equation~\eqref{eq7}) for fundamental radial in-plane axisymmetric vibration mode with free edge boundary condition is equal to $2.0424$.
The PZT transducer used in our experiment is square-shaped with edge length \SI{20}{\mm}, that therefore corresponds to an equivalent circular PZT transducer with radius \SI{13}{\mm}, as shown in Figure~\ref{fig:eqv_circ}. 

\subsection{Mathematical formulation of bi-layer circular plate with equivalent circular PZT transducer}

\begin{figure}[!bp]
    \centering
    \includegraphics[width = 0.7\linewidth]{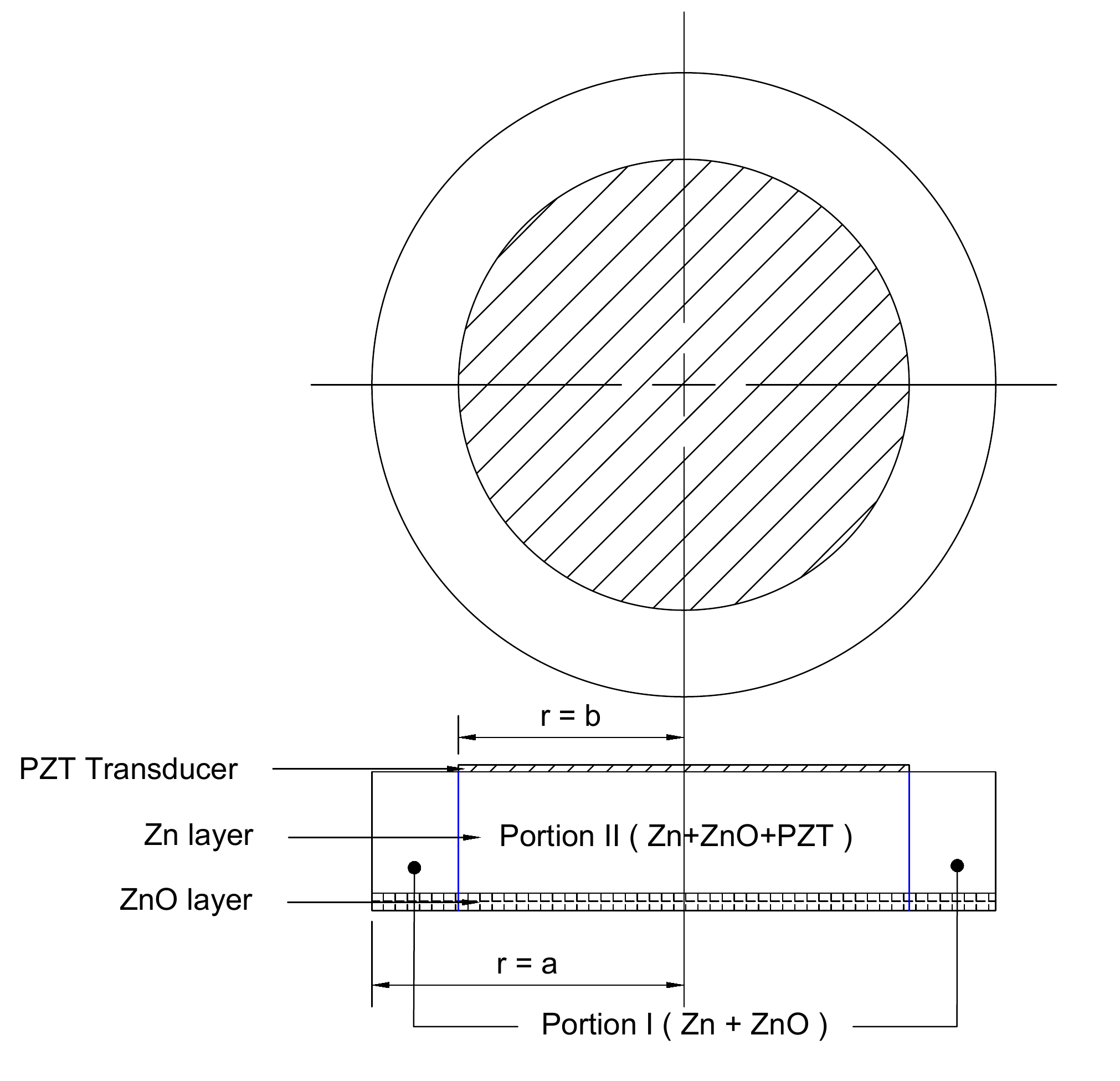}
    \caption{(top) Top view and (bottom) cross-section of the circular zinc anode with equivalent circular PZT transducer on top surface and uniformly formed zinc oxide layer on bottom surface.}
    \label{fig:bilayer}
\end{figure}

To derive analytical expression for the resonance frequency of fundamental axisymmetric radial expansion mode of the corroded zinc plate with the PZT transducer, we model the structure as a composite of two portions shown in Figure~\ref{fig:bilayer}.
The layers are considered perfectly `bonded' with each other i.e. the layers have equal radial displacements and there is no inter-layer slip. This assumption implies there is no delamination between the layers, and is consistent with experimental observations during initial corrosion \cite{meng2019initial}.
Thus, the circular zinc anode with thin corroded layer (zinc oxide) can be divided into two portions namely: Portion I (two layers: zinc + zinc oxide), and Portion II (three layers: zinc + zinc oxide + equivalent circular PZT).
The following assumptions are considered for the analysis:
\begin{itemize}
  \item All materials are assumed to be isotropic
  \item Corrosion occurs uniformly across the bottom surface of the zinc anode
\end{itemize}

Therefore, the general solutions for axisymmetric in-plane radial expansion of the circular disc in the two portions are expressed as:

\begin{equation}
U(r) = \begin{cases}
U_1(r)=C_1 J_1\left(\beta_1\frac{r}{a}\right)+ C_2 Y_1\left(\beta_1\frac{r}{a}\right) &\text{Portion I}\\
U_2(r)=C_3 J_1\left(\beta_2\frac{r}{a}\right) &\text{Portion II}
\end{cases}
\end{equation}
where the dimensionless frequency parameters are given as (subscripts $1$, $2$ and $3$ denote zinc oxide layer, zinc layer and PZT transducer, respectively):

\begin{equation}
    {\beta_1}=\sqrt{\frac{\left(\rho_1 h_1+\rho_2 h_2\right)a^2}{\frac{E_1 h_1}{1-\nu_1^2}+\frac{E_2 h_2}{1-\nu_2^2}}}\times\omega
\end{equation}
\begin{equation}
    {\beta_2}=\sqrt{\frac{\left(\rho_1 h_1+\rho_2 h_2+\rho_3 h_3\right)a^2}{\frac{E_1 h_1}{1-\nu_1^2}+\frac{E_2 h_2}{1-\nu_2^2}+\frac{E_3 h_3}{1-\nu_3^2}}}\times\omega
\end{equation}
where $h$ is the thickness, $E$ is the Young's modulus, $\rho$ is the density and $\nu$ is the Poisson's ratio of the respective layer. The Young's modulus for the PZT transducer is obtained from the compliance coefficients as $E_3=1/S_{11}$ and the Poisson's ratio as $\nu_3=-S_{12}/S_{11}$, wherein the compliance coefficients $S_{11}$ and $S_{12}$ are obtained from the compliance matrix of the PZT transducer.
The radial force expression for Portion I and Portion II can be expressed as: 

\begin{equation}
N_r = \begin{cases}
N_{r1}=D_1\left(\frac{\partial U_1}{\partial r}+\nu_{1eq}\frac{U_1}{r}\right) &\text{Portion I}\\
N_{r2}=D_2\left(\frac{\partial U_2}{\partial r}+\nu_{2eq}\frac{U_2}{r}\right) &\text{Portion II}
\end{cases}
\end{equation}
where the stiffness coefficients $D_1$ and $D_2$, and the equivalent Poisson's ratio $\nu_{1eq}$ amd $\nu_{1eq}$ for Portions I and II respectively are expressed as follows:
\begin{equation}
    D_1=\frac{E_1 h_1}{1-\nu_1^2}+\frac{E_2 h_2}{1-\nu_2^2}
\end{equation}

\begin{equation}
    D_2=\frac{E_1 h_1}{1-\nu_1^2}+\frac{E_2 h_2}{1-\nu_2^2}+\frac{E_3 h_3}{1-\nu_3^2}
\end{equation}

\begin{equation}
    \nu_{1eq}=\frac{\frac{\nu_1 E_1 h_1}{1-\nu_1^2}+\frac{\nu_2 E_2 h_2}{1-\nu_2^2}}{\frac{E_1 h_1}{1-\nu_1^2}+\frac{E_2 h_2}{1-\nu_2^2}}
\end{equation}
\begin{equation}
    \nu_{2eq}=\frac{\frac{\nu_1 E_1 h_1}{1-\nu_1^2}+\frac{\nu_2 E_2 h_2}{1-\nu_2^2}+\frac{\nu_3 E_3 h_3}{1-\nu_3^2}}{\frac{E_1 h_1}{1-\nu_1^2}+\frac{E_2 h_2}{1-\nu_2^2}+\frac{E_3 h_3}{1-\nu_3^2}}
\end{equation}

\subsection{Determination of resonance frequency of fundamental radial expansion mode}
\label{analytical_calculation}
For the two layer (zinc and zinc oxide) circular plate with equivalent circular PZT patch undergoing axisymmetric in-plane radial vibration, the free edge boundary condition at $r=a$ is $N_{r1}=0$ (zero radial force) i.e. $\frac{\partial U_1}{\partial r}\Big|_{r=a}+\nu_{1eq}\frac{U_1}{r}\Big|_{r=a}=0$.
At the interface between Portion I and Portion II at $\textit{r=b}$, there exists continuity of radial displacement and radial forces, i.e. $U_1(r=b)=U_2(r=b)$ and $N_{r1}\Big|_{r=b}=N_{r2}\Big|_{r=b}$.
Substituting the expressions for displacements in the above equations, a $3\times3$ characteristic determinant is obtained as below:

\begin{equation}
    \Delta=
    \left|
\begin{array}{ccc}
    K_{11} & K_{12} & K_{13} \\
    K_{21} & K_{22} & K_{23} \\
    K_{31} & K_{32} & K_{33} \\
\end{array}
\right|
\end{equation}
where the elements of the determinant are:
\begin{equation}
    K_{11}=\beta_1 J_0\left(\beta_1\right)-J_1\left(\beta_1\right) \left(1-\nu_{1eq}\right)
\end{equation}
\begin{equation}
    K_{12}=\beta_1 Y_0\left(\beta_1\right)-Y_1\left(\beta_1\right) \left(1-\nu_{1eq}\right)
\end{equation}
\begin{equation}
    K_{13}=0
\end{equation}
\begin{equation}
    K_{21}=J_1\left(\frac{b \beta_1}{a}\right)
\end{equation}
\begin{equation}
    K_{22}=Y_1\left(\frac{b \beta_1}{a}\right)
\end{equation}
\begin{equation}
    K_{23}=-J_1\left(\frac{b \beta _2}{a}\right)
\end{equation}
\begin{equation}
    K_{31}=\frac{\beta _1 J_0\left(\frac{b \beta_1}{a}\right)}{a}-\frac{\left(1-\nu_{1eq}\right) J_1\left(\frac{b \beta_1}{a}\right)}{b}
\end{equation}
\begin{equation}
    K_{32}=\frac{\beta_1 Y_0\left(\frac{b \beta_1}{a}\right)}{a}-\frac{\left(1-\nu_{1eq}\right) Y_1\left(\frac{b \beta_1}{a}\right)}{b}
\end{equation}
\begin{equation}
    K_{33}=\frac{D_2 \left(-\frac{\beta_2 J_0\left(\frac{b \beta_2}{a}\right)}{a}+\frac{\left(1-\nu_{2eq}\right) J_1\left(\frac{b \beta_2}{a}\right)}{b}\right)}{D_1}
\end{equation}

Since the material properties ($E$, $\nu$ and $\rho$) are known for the three layers, the characteristic determinant $\Delta$ can be expressed as a function of the natural frequency $\omega$, and thickness of the corroded zinc oxide layer $h_1$ and the uncorroded zinc layer $h_2$, i.e. $\Delta=\Delta(\omega,h_1,h_2)$. The natural frequencies of the system undergoing corrosion can be obtained by setting the characteristic determinant to zero. The thicknesses $h_1$ and $h_2$ are obtained using Faraday's law of electrolysis. Thus with the values of $h_1$ and $h_2$ determined, the characteristic equation can be expressed as $\Delta(\omega)=0$, solving which gives the natural frequency of the system. The characteristic equation is a transcendental equation that is solved using MATLAB.

\subsection{Comparison of solutions of analytical model and FEA}\label{comparison}

The solution for  resonance frequency of fundamental axisymmetric radial expansion mode obtained from the analytical model described above are baselined against FEM model of the structure solved using COMSOL Multiphysics. We solve for a commercially available cylindrical zinc sacrificial anode (Canode, Krishna Conchem Products Pvt. Ltd.) of diameter \SI{3.6}{\cm} and thickness \SI{0.73}{\cm}, and PZT-5H transducer of dimensions \SI{20}{}$\times$\SI{20}{}$\times$\SI{0.4}{\mm^3} (SP-5H, Sparkler Ceramics Pvt. Ltd.). A photograph of the assembly, and the FEM model is shown in Figure~\ref{fig:structure_mode}.
The equivalence of the two structures shown in Figure~\ref{fig:eqv_circ} is validated through Finite Element (FE) simulations performed using COMSOL Multiphysics, which yields resonance frequencies for the fundamental axisymmetric radial expansion mode as \SI{69.42}{kHz} and \SI{69.44}{kHz} (amounting to $<$\SI{0.03}{\%} error) for the circular zinc plate with square-shaped PZT and equivalent circular PZT, respectively.
The FE simulations are performed using ``Fine'' mesh setting in COMSOL Multiphysics, considering the trade-off between computation time and convergence of the eigenfrequency solution (as shown in Table~\ref{tbl:mesh}. The computation time is obtained on a computer with $64$-bit Intel\textregistered~Xeon\textregistered~E5-1650 v4 \SI{3.60}{GHz} CPU and \SI{32}{GB} RAM. The computation time is significantly higher upon introduction of zinc oxide layer as described below.

\begin{table}[!tbp]
\begin{center}
\caption{Comparison of impact of mesh settings in COMSOL Multiphysics on eigenfrequency solution}
\label{tbl:mesh} 
\begin{tabular}{c|c|c|c} 
\hline
\multicolumn{1}{p{3cm}}{\centering Mesh setting} &
\multicolumn{1}{|p{3cm}|}{\centering Number of domain elements} &
\multicolumn{1}{p{2cm}}{\centering Computation time [\SI{}{\second}]}  & 
\multicolumn{1}{|p{2cm}}{\centering Eigenfrequency [\SI{}{\kilo\Hz}]} \\[0.5ex]
\hline
Extremely Coarse  & 438 & 3 & 69.471\\[0.5ex]
Extra Coarse  & 2079 & 3 & 69.431\\[0.5ex]
Coarser  & 4522 & 4 & 69.425\\[0.5ex]
Coarse  & 10070 & 8 & 69.422\\[0.5ex]
Normal  & 18985 & 13 & 69.421\\[0.5ex]
Fine & 29686 & 21 & 69.42\\[0.5ex]
Finer  & 48591 & 36 & 69.42\\[0.5ex]
\hline
\end{tabular}
\end{center}

\end{table}

When corrosive agents attack the sacrificial anode, a thin layer of corrosion products forms on its surface due to the multitude of corrosive reactions taking place. Thus for simplicity we have assumed that soon after corrosion initiates the anode is part metal and part metal oxide \cite{meng2019initial,perkins1977anodic}. The surface of the anode on which the PZT is attached is assumed to not corrode (this is ensured using water-proofing sealant in the experiment, as described in the following section).
In the initial stages of accelerated corrosion due to externally applied impressed current, zinc reacts with oxygen and forms zinc oxide.
By calculating the amount of mass (thickness) loss in the zinc layer and added mass (thickness) due to formation of zinc oxide using Faraday's law of electrolysis, we obtain the resonance frequency for the radial expansion mode through FE simulations performed in COMSOL Multiphysics and using the analytical method described in section~\ref{analytical_calculation}. Table~\ref{tbl:properties} lists the material properties for zinc and zinc oxide used for these calculations.

\begin{table}[!bp]
\begin{center}
\caption{Material properties used for calculating mass conversion of zinc to zinc oxide}
\label{tbl:properties} 
\begin{tabular}{c|c|c} 
\hline
\multicolumn{1}{p{2cm}|}{\centering Material} & 
\multicolumn{1}{p{3cm}}{\centering Density [\SI{}{\gram\per\cm^3}]} &
\multicolumn{1}{|p{4cm}}{\centering Molecular mass [\SI{}{\gram\per\mole}]}
\\[0.5ex]
\hline
Zinc & 7.14 & 65.38 \\[0.5ex]
Zinc oxide & 5.68 & 81.38 \\[0.5ex]
\hline
\end{tabular}
\end{center}
\end{table}

The conversion of zinc to zinc oxide is governed by the chemical reaction:
\begin{equation}
\ce{Zn + 0.5O_2\xrightarrow{}ZnO}
\end{equation}

\begin{figure}[!tbp]
    \centering
    \includegraphics[width = 0.5\linewidth ]{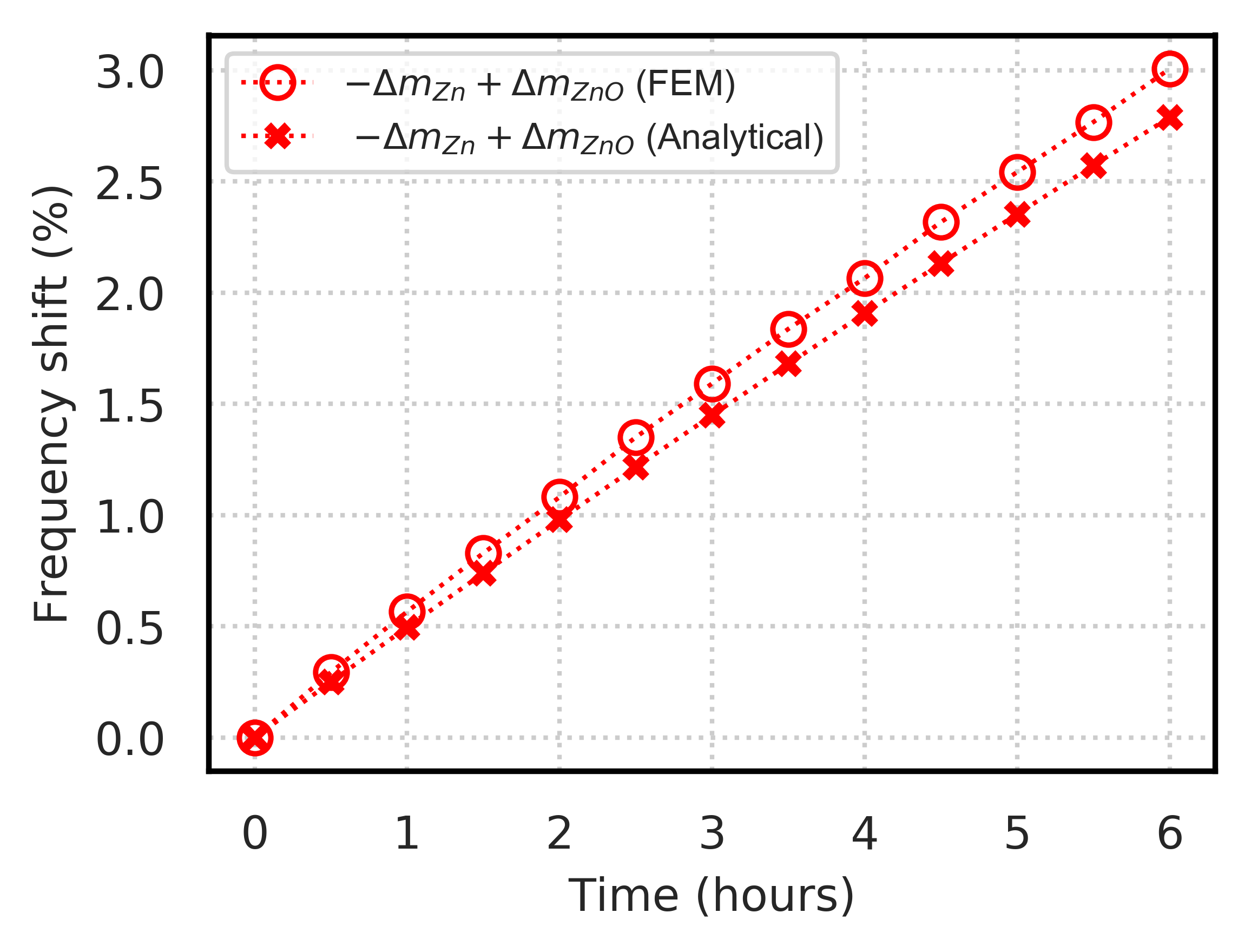}
    \caption{Frequency shift of the radial mode of a zinc anode of diameter \SI{3.6}{\cm} and thickness \SI{0.73}{\cm}, and PZT-5H transducer of dimensions \SI{20}{}$\times$\SI{20}{}$\times$\SI{0.4}{\mm^3}, due to \SI{0.35}{A} impressed current applied to induce accelerated corrosion. The shift in resonance frequency relative to the value at $t=$ \SI{0}{\hour} (commencement of impressed current application) is shown at intervals of \SI{30}{\minute} up to $t=$ \SI{6}{\hour}.
    The results show good agreement of the analytical model with Finite Element Method (FEM).}
    \label{fig:comsol_ana_f_shift}
\end{figure}

Thus, \SI{1}{\gram} of zinc produces \SI{1.24}{\gram} of zinc oxide i.e. \SI{1}{\cm^3} volume of zinc is converted to \SI{1.57}{\cm^3} of zinc oxide. Since we consider a cylindrical geometry whose radius does not change due to corrosion (i.e. zinc oxide only forms on the bottom circular face of the anode due to waterproofing of top and side walls of the disc in the experiment), reduction of thickness of the zinc layer by \SI{1}{\cm} results in formation of \SI{1.57}{\cm} of zinc oxide. The corresponding thickness reduction of zinc due to \SI{0.35}{\A} impressed current applied for various time duration values shown in Table~\ref{tbl:bins2} are obtained using Faraday's law of electrolysis. Figure~\ref{fig:comsol_ana_f_shift} shows a comparison of the FEM and analytical model results accounting for Faradaic mass loss of zinc $(\Delta m_{\ce{Zn}})$, and added mass of the zinc oxide layer formed due to corrosion $(\Delta m_{\ce{ZnO}})$.

\begin{table}[!tbp]
\begin{center}
\caption{Calculated thickness loss of zinc layer and thickness added in zinc oxide layer for \SI{0.35}{\A} impressed current applied for various time duration values. The resonance frequencies obtained using FEM simulation and analytical calculations show good agreement with each other.}
\label{tbl:bins2} 
\begin{tabular}{c|c|c|c|c} 
\hline
\multicolumn{1}{p{1cm}}{\centering Time [\SI{}{\minute}]} & 
\multicolumn{1}{|p{2.5cm}}{\centering \ce{Zn} thickness loss [\SI{}{\cm}]} &
\multicolumn{1}{|p{2.5cm}|}{\centering \ce{ZnO} thickness gain [\SI{}{\cm}]} &
\multicolumn{1}{p{2cm}}{\centering Analytical result [\SI{}{\kilo\Hz}]}  & 
\multicolumn{1}{|p{2cm}}{\centering FEM result [\SI{}{\kilo\Hz}]} \\[0.5ex]
\hline
0   &0 &0  &69.875  &69.42 \\[0.5ex]
30  &0.0029 &0.0045 &70.194  &69.717 
 \\[0.5ex]
60  &0.0058 &0.0091 &70.510  &70.009 \\[0.5ex]
90  &0.0088 &0.0137 &70.825  &70.299 \\[0.5ex]
120 &0.0117 &0.0183 &71.137  &70.585 \\[0.5ex]
150 &0.0146 &0.0229 &71.447  &70.869 \\[0.5ex]
180 &0.0176 &0.0275 &71.756  &71.149 \\[0.5ex]
210 &0.0205 &0.0321 &72.062  &71.427 \\[0.5ex]
240 &0.0234 &0.0367 &72.367  &71.703 \\[0.5ex]
270 &0.0264 &0.0413 &72.669  &71.975 \\[0.5ex]
300 &0.0293 &0.0459 &72.970  &72.246 \\[0.5ex]
330 &0.0323 &0.0505 &73.269  &72.514 \\[0.5ex]
360 &0.0352 &0.0551 &73.567  &72.78 \\[0.5ex]
\hline
\end{tabular}
\end{center}

\end{table}

\section{Experimental Setup}
\begin{figure}[!tbp]
    \centering
    \includegraphics[width = 1\linewidth ]{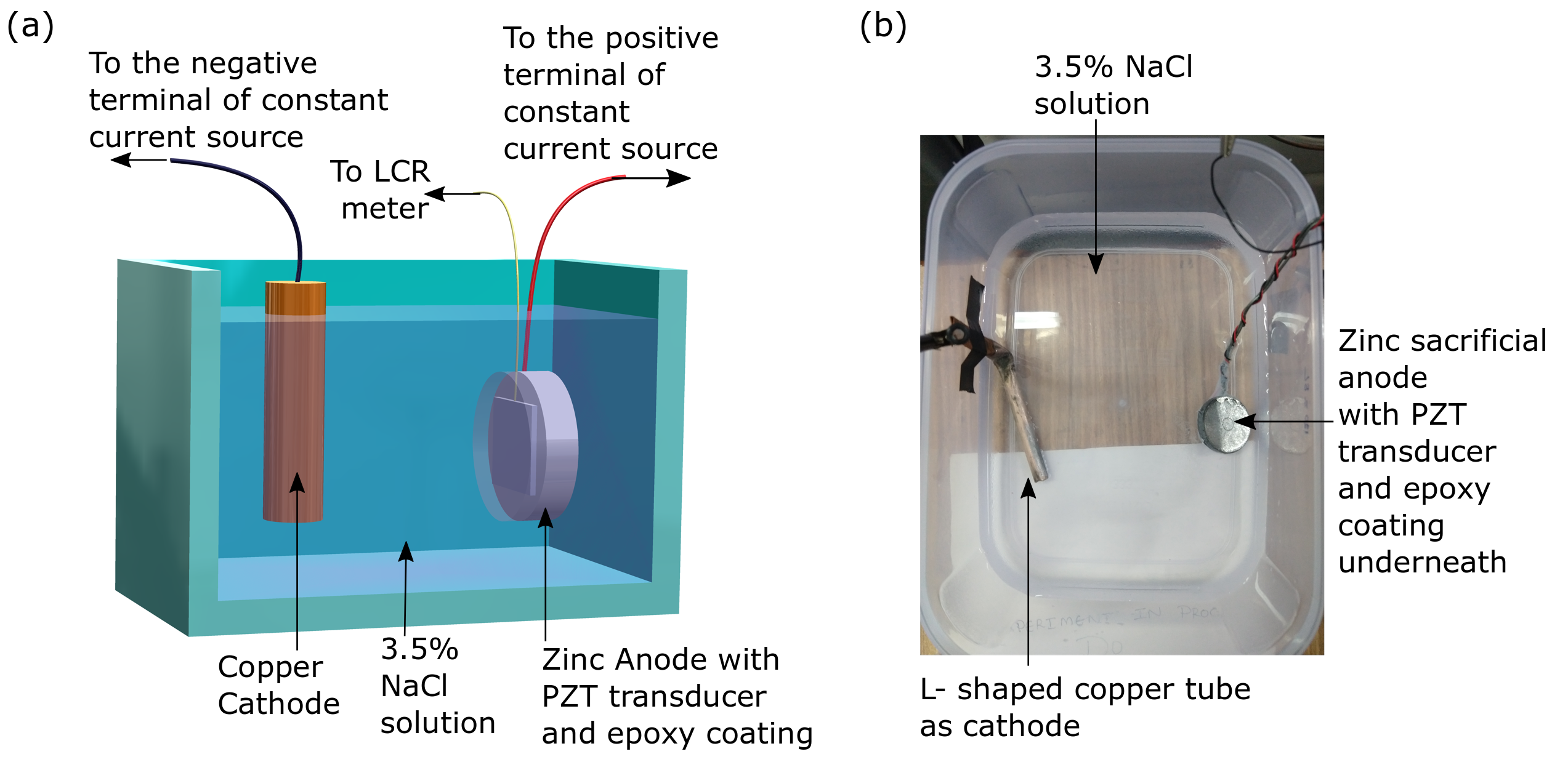}
    \caption{(a) Illustration and (b) photograph of the experimental setup used for accelerated corrosion of sacrificial anode.}
    \label{fig:setup}
\end{figure}

Two samples (A and B) of the sacrificial anode with PZT transducer as described in section~\ref{comparison} are fabricated. One of the circular surfaces of the zinc anode is gently polished with fine grit sandpaper and the PZT transducer is affixed on the center of the polished side with Fewikwik\textregistered~instant adhesive.
Further, a coating of waterproofing epoxy (M-Seal\textregistered~Clear RTV Silicone Sealant) is applied over the PZT transducer to protect it from the liquid electrolyte used in the experiment. This waterproofing sealant also protects the polished surface of the anode from corroding. The assembly is left undisturbed for \SI{12}{\hour} to allow the epoxy to cure. 
The experimental setup for accelerated corrosion by means of applied impressed current (shown in Figure \ref{fig:setup}) consists of an electrolytic cell with a set of two electrodes submerged in an electrolyte. We use a copper tube as cathode and a \SI{3.5}{\%} \ce{NaCl} solution as electrolyte. The copper electrode is connected to negative terminal of a constant current source (Aplab LQ6324T) while positive terminal of the current source is connected to the zinc anode assembly.
The impedance of the PZT transducer is measured using a precision LCR meter (Agilent E4980A). The LCR meter is connected to a computer via USB, and is controlled using a custom-made software designed with LabVIEW for data acquisition.
The constant current source supplies current of magnitude \SI{0.35}{\A}. After every \SI{30}{\minute} interval, the current source is turned off and impedance of the PZT transducer is recorded, without disturbing the mechanical arrangement of the apparatus.

\section{Experimental Results and Discussion}
\begin{figure}[!tbp]
    \centering
    \includegraphics[width = 
    \linewidth ]{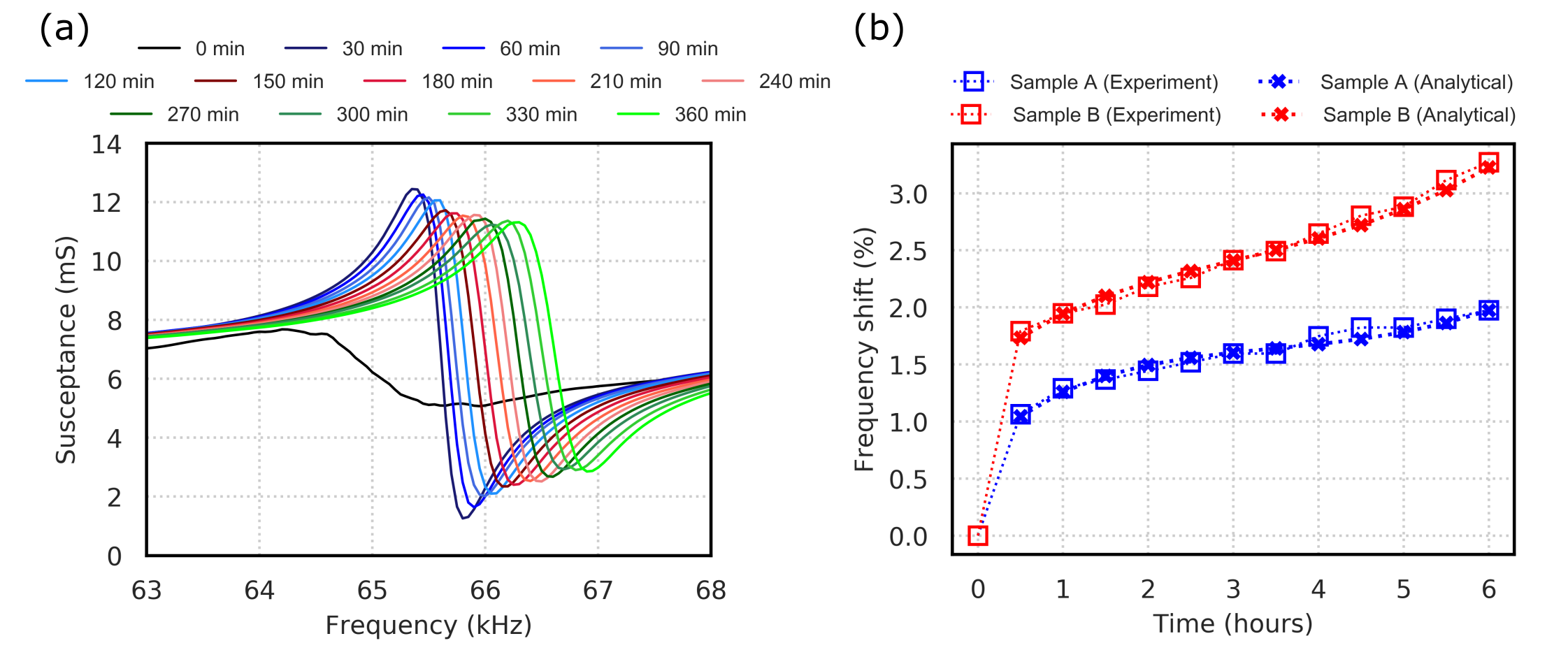}
    \caption{(a) Evolution of susceptance spectra (for sample B) of the PZT transducer on sacrificial anode system. The series resonance frequency (corresponding to peak in the susecptance spectra) changes with increasing corrosion of the anode and is used as a signature for estimating extent of corrosion.
    (b) Experimentally measured shift in resonance frequency of radial expansion mode for samples A and B compared with results obtained from analytical model (equation~\eqref{gamma_values}).}
   \label{fig:results}
\end{figure}

Figure~\ref{fig:results} shows experimentally measured change in series resonance frequency of the susceptance spectra over duration of the experiment. Both samples show a change in slope and nonlinearity in the measurements (as indicated by the blue and red hollow squares in Figure~\ref{fig:results}(b)) instead of the linear trend as predicted by Faraday's law (Figure~\ref{fig:comsol_ana_f_shift}). 
The assumption made in Figure~\ref{fig:comsol_ana_f_shift} is that corrosion products stay on the anode throughout the duration of the experiment. While performing the experiments, we observe that residue of the corrosion product separates from the anode into the electrolyte.
The mathematical model of the process of delamination is explained below.
The ideal mass-loss equation as depicted by Faraday's law is given by:

\begin{equation}\label{Faradaylaw}
\Delta m=\frac{I M t}{F z}
\end{equation}
where $\Delta m$ is the mass loss, $I$ is current passing through the cell, $M$ is molecular weight of the reactive species, $t$ is total time duration for which current flows in the cell, $F$ is Faraday's constant, and $z$ is valency of the reactive species. 
As per Faraday's law, the mass of the sacrificial anode assembly (i.e. circular zinc anode with affixed PZT transducer and waterproofing epoxy layer) should increase as time progresses, due to formation of the zinc oxide layer. However, we observe a reduction in overall mass of the sacrificial anode assembly in the experiment (\SI{2.51}{\gram} and \SI{2.14}{\gram} for samples A and B respectively, measured with a precision weighing scale). The reduction in overall mass is attributed to partial delamination of zinc oxide from the anode surface. 
The change in delamination front of a polymeric coating from zinc surface in the presence of electrolyte shows a power law dependence with time \cite{furbeth}:
\begin{equation}\label{adel}
    a_{del}=k \times t^A
\end{equation}
where $a_{del}$ is the distance of the delamination front from the edge of the plate, $k$ is the rate constant, $t$ is the exposure time for delamination to occur, and $A$ is the exponent in the power law.
Thus, the volume of the non-delaminated zinc oxide layer can be expressed as:
\begin{equation}
V_{\text{non-delaminated}}=\pi\left(a-a_{\text {del}}\right)^2h_1
\end{equation}

To simplify modeling of the disc, we assume that reduction in volume of the zinc oxide layer happens only due to the reduction of thickness and that the area remains the same. This assumption is consistent with the experimental observation that corrosion occurs uniformly over the surface and no delamination is observed between the formed zinc oxide layer and the zinc layer in the initial \SI{30}{\minute} phase of the experiment {\cite{meng2019initial}}. Thus, the reduction of overall mass of the sacrificial anode assembly happens from the dissolution or the delamination of the Zinc oxide layer.
Therefore we define a parameter termed as the non-delaminated factor $\Gamma_d$, which is the ratio of the non-delaminated thickness of zinc oxide layer $h_{1d}$ to the total thickness of the zinc oxide layer in absence of delamination $h_{1}$:
\begin{equation}\label{gamma}
\Gamma_{d}=\frac{h_{1d}}{h_1}=\frac{V_{\text {non-delaminated}}}{V}=\frac{\pi\left(a-a_{\text {del}}\right)^{2}}{\pi a^{2}} = 1 - \frac{2a_{del}}{a} + \frac{a_{del}^2}{a^2}
\end{equation}

Using equations \eqref{adel} and \eqref{gamma}, we can generalize the non-delamination factor for a circular disc undergoing corrosion as:

\begin{equation}\Gamma_{d}=1-C_{1} t^{A}+C_{2} t^{2 A}\end{equation}
where $C_{1}$ and $C_{2}$ are reaction dependent constants. According to the theory of rate-determining step, the value of the exponent $A$ may vary between \SI{0.5}{} and \SI{1.0}{} \cite{furbeth}.
The constants $C_{1}$, $C_{2}$ and $A$ are estimated for both Sample A and Sample B using MATLAB Curve Fitting Toolbox,
and the non-delamination factors for Sample A and Sample B thus obtained are:
\begin{equation}\label{gamma_values}
\Gamma_d = \begin{cases}
1-0.4676t^{0.695}+0.06953t^{2\times0.695} &\text{Sample A}\\
1-0.4925t^{0.65}+0.08753t^{2\times0.65} &\text{Sample B}
\end{cases}
\end{equation}

\begin{figure}[!tbp]
    \centering
    \includegraphics[width = \linewidth ]{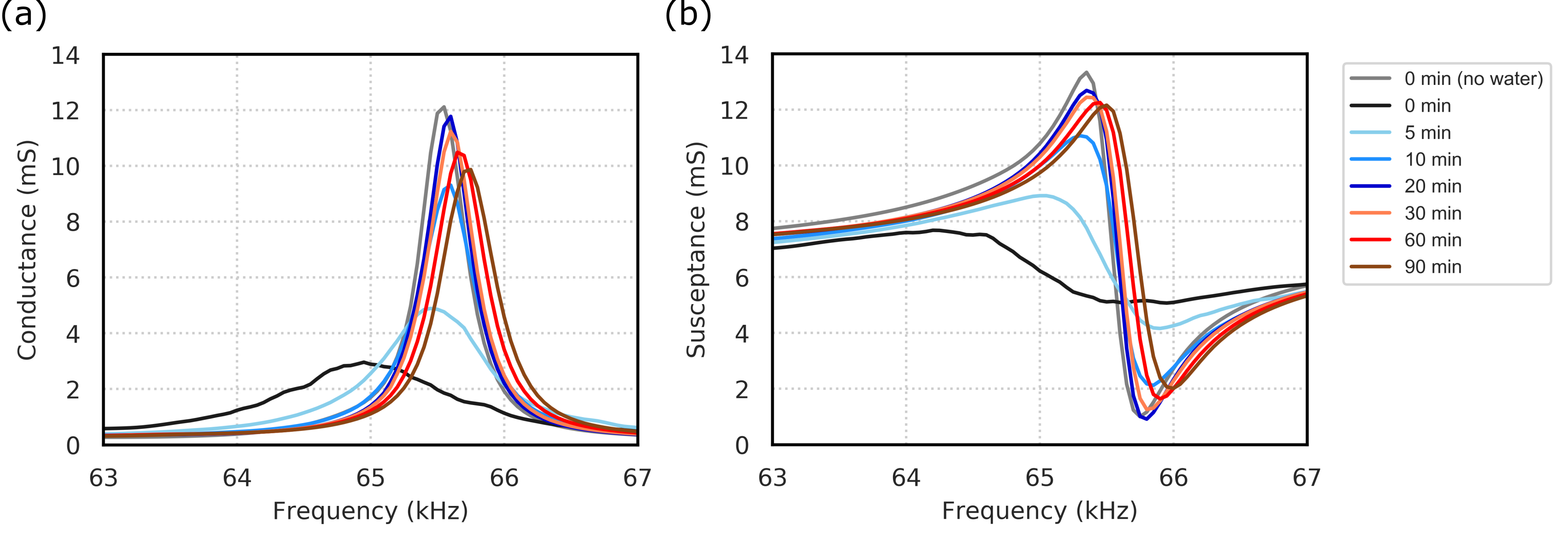}
    \caption{(a) Conductance, and (b) susceptance spectra for the sacrificial anode + PZT assembly (sample B) showing measurements obtained up to $t=$ \SI{90}{\minute}. The resonance frequency changes drastically in the first few minutes of the experiment, and at $t=$ \SI{20}{\minute} the resonance frequency attains a value close to the case where the uncorroded anode is not submerged in electrolyte (labeled as `0 min (no water)'). Additionally we also see abrupt change in the loading of the resonance (i.e. reduction of peak height and broadening of peak width) upon submerging the assembly in electrolyte (labeled `0 min'). The loading reduces with progressing time. The origins of this observation are under investigation.}
   \label{fig:initial_shift}
\end{figure}

The overall mass change of the sacrificial anode can be attributed to the mass loss of zinc 
($-\Delta m_{\ce{Zn}}$) due to electrolysis, and mass gain due to the non-delaminated portion of zinc oxide ($+\Gamma_{d} \times \Delta m_{\ce{ZnO}}$): 
\begin{equation}
    \text{Overall mass loss}=-\Delta m_{\ce{Zn}}+\Gamma_d \times \Delta m_{\ce{ZnO}}
\end{equation}
The estimated overall mass loss calculated using the constants obtained from the curve fitting are \SI{1.87}{\gram} and \SI{1.54}{\gram} for sample A and sample B , respectively.
We observe a large frequency shift from $t=$ \SI{0}{\hour} to $t=$ \SI{0.5}{\hour} for both samples, that is not captured by $\Gamma_d$.
In order to account for this large initial shift in resonance frequency, we have to introduce an increment of \SI{0.714}{\%} in the resonance frequency at $t=$ \SI{0}{\hour} for Sample A and \SI{1.41}{\%} for Sample B for best least squares fit.
The dramatic variation in the frequency shift in the first \SI{30}{\minute} of the experiment is shown in the conductance and susceptance spectra of sample B in Figures~\ref{fig:initial_shift}(a) and (b), respectively.
The resonance frequency changes drastically in the first few minutes of the experiment. At $t=$ \SI{20}{\minute}, the resonance frequency for the anode assembly submerged in the electrolyte attains a value close to the case where the uncorroded anode is not submerged in the electrolyte. In addition to the sharp increase in resonance frequency, we also notice significant loading of the resonance (reduction of peak height and broadening of peak width) at $t=$ \SI{0}{\minute} upon submerging the anode in electrolyte. With progressing time, the impedance spectra show reduction in loading and the subsequent frequency shift is explained by the delamination model described above. The origin for the drastic change in loading and initial frequency shift of the radial expansion mode of the circular zinc disc anode is under investigation.

\section{Conclusion}

In summary, we have demonstrated an electromechanical impedance based method to detect early corrosion of zinc sacrficial anodes (used in cathodic 
protection systems). 
The experiments are conducted by connecting the sacrificial anode assembly in an electrochemical half-cell comprising of copper cathode and \SI{3.5}{\%} \ce{NaCl} electrolyte solution. A constant current source is used to generate impressed current for accelerated corrosion of the anode, and a precision LCR meter is used to measure the electro-mechanical impedance of the PZT transducer. The electro-mechanical resonance frequency is obtained from the susceptance spectra recorded using a LabVIEW program running on a computer connected to the LCR meter.
We have chosen a commercially available circular zinc plate for our experiments, with PZT-5H transducer affixed in the center of one of the circular faces of the plate. We present an analytical model to calculate change in resonance frequency of the radial expansion vibration mode of the assembly due to corrosion of zinc and resultant formation of zinc oxide.

To the best of our knowledge, this is the first reported realization of a smart-cathodic protection system with built-in sensing of extent of degradation of the sacrificial anode. The analytical model presented in our work accounts for nonlinearity due to partial delamination of the zinc oxide film, as was observed in our experiments conducted with two such sacrificial anode samples.
The analytical model and FEA framework presented in this work can be adapted to study the variation of any desired bulk acoustic vibration modes of sacrificial anodes of any arbitrary geometry.
Future work will focus on understanding the origins of the sharp change observed in resonance frequency in first \SI{30}{\minute} of the experiment as well as drastic change in loading of the resonance.
The technique proposed in this paper can be used for in-situ monitoring of sacrificial anodes and detecting early signs of cover deterioration due to corrosive agents in marine infrastructure prone to corrosion. This technique is cost-effective and can be equipped with portable impedance analyzers (e.g. Analog Devices AD5933) and remote monitoring capabilities and thus has a wide application potential in structural health monitoring. This technique could also be adapted for real-time monitoring of degradation of electrodes in batteries and fuel cells.

\section*{CRediT authorship contribution statement}
\textbf{Durgesh Tamhane:} Conceptualization, Methodology, Investigation, Software, Validation, Visualization, Writing - Original Draft.
\textbf{Jeslin Thalapil:} Formal analysis, Investigation, Software, Validation, Visualization, Writing - Original Draft.
\textbf{Sauvik Banerjee:} Conceptualization, Methodology, Investigation, Resources, Funding acquisition, Supervision, Writing - Original Draft, Writing - Review \& Editing.
\textbf{Siddharth Tallur:} Conceptualization, Methodology, Investigation, Resources, Funding acquisition, Supervision, Writing - Original Draft, Writing - Review \& Editing.

\section*{Declaration of Competing Interest}
The authors declare that they have no known competing financial interests or personal relationships that could have appeared to influence the work reported in this paper.

\section*{Acknowledgments}
This work was supported by IMPRINT-2A by Department of Science \& Technology, Government of India [grant IMP/2018/001442] with additional financial support from Sanrachana Structural Strengthening Pvt. Ltd. [grant RD/0119-SSIMPQ2-001]. The authors thank Krishna Conchem Products Pvt. Ltd. for providing the zinc sacrificial anodes used for performing the experiments. The authors also thank Mr. Hrishikesh Belatikar at IIT Bombay for assistance with developing the LabVIEW based data acquisition software, Dr. Rajeshwara Chary Sriramadasu at IIT Bombay for preliminary discussions, and Wadhwani Electronics Lab (WEL) at IIT Bombay for providing Agilent E4980A precision LCR meter.

\appendix
\section{General solution for axisymmetric in-plane radial vibration of a circular plate}
\label{app:a}

In case of axisymmetric radial vibration of a circular plate, the displacement $u_r$ is independent of the angular position $\theta$ i.e. $\frac{\partial u_r}{\partial \theta}=0$. Additionally, tangential displacement $u_\theta=0$, since motion is only along the radial direction. The mode shape of the radial expansion mode is shown as an inset in Figure~\ref{fig:structure_mode}(c).
Thus, strain-displacement relations for axisymmetric motion can be expressed as:
\begin{equation}\label{eq:A1}
    \epsilon_r=\frac{\partial u_r}{\partial r}
\end{equation}
\begin{equation}
    \epsilon_\theta=\frac{u_r}{r}
\end{equation}

The stress-strain relationship for case of plane stress in cylindrical coordinates can be expressed as:
\begin{equation}
    \sigma_r=\frac{E}{1-\nu^2}\left(\epsilon_r+\nu \epsilon_\theta \right)
\end{equation}
\begin{equation}
    \sigma_\theta=\frac{E}{1-\nu^2}\left(\nu \epsilon_r+\epsilon_\theta \right)
\end{equation}
\begin{equation}\label{eq:A5}
    \sigma_{r\theta}=0
\end{equation}
where $E$ and $\nu$ denote Young's modulus and Poisson's ratio for the material of the plate, respectively.
Using equations~\eqref{eq:A1}$-$\eqref{eq:A5}, we can express stress-displacement relations for case of axisymmetric plane stress as:
\begin{equation}
    \sigma_r=\frac{E}{1-\nu^2}\left(\frac{\partial u_r}{\partial r}+\nu \frac{u_r}{r} \right)
\end{equation}
\begin{equation}
    \sigma_\theta=\frac{E}{1-\nu^2}\left(\nu \frac{\partial u_r}{\partial r}+\frac{u_r}{r} \right)
\end{equation}

Further, we can express radial and circumferential forces in the plate as:
\begin{equation}\label{eq:A8}
    N_r=\frac{E h}{1-\nu^2}\left(\frac{\partial u_r}{\partial r}+\nu \frac{u_r}{r} \right)
\end{equation}
\begin{equation}\label{eq:A9}
    N_\theta=\frac{E h}{1-\nu^2}\left(\nu \frac{\partial u_r}{\partial r}+\frac{u_r}{r} \right)
\end{equation}
\begin{equation}
    N_{r\theta}=0
\end{equation}

\begin{figure}[!tbp]
    \centering
    \includegraphics[width = 0.9\linewidth ]{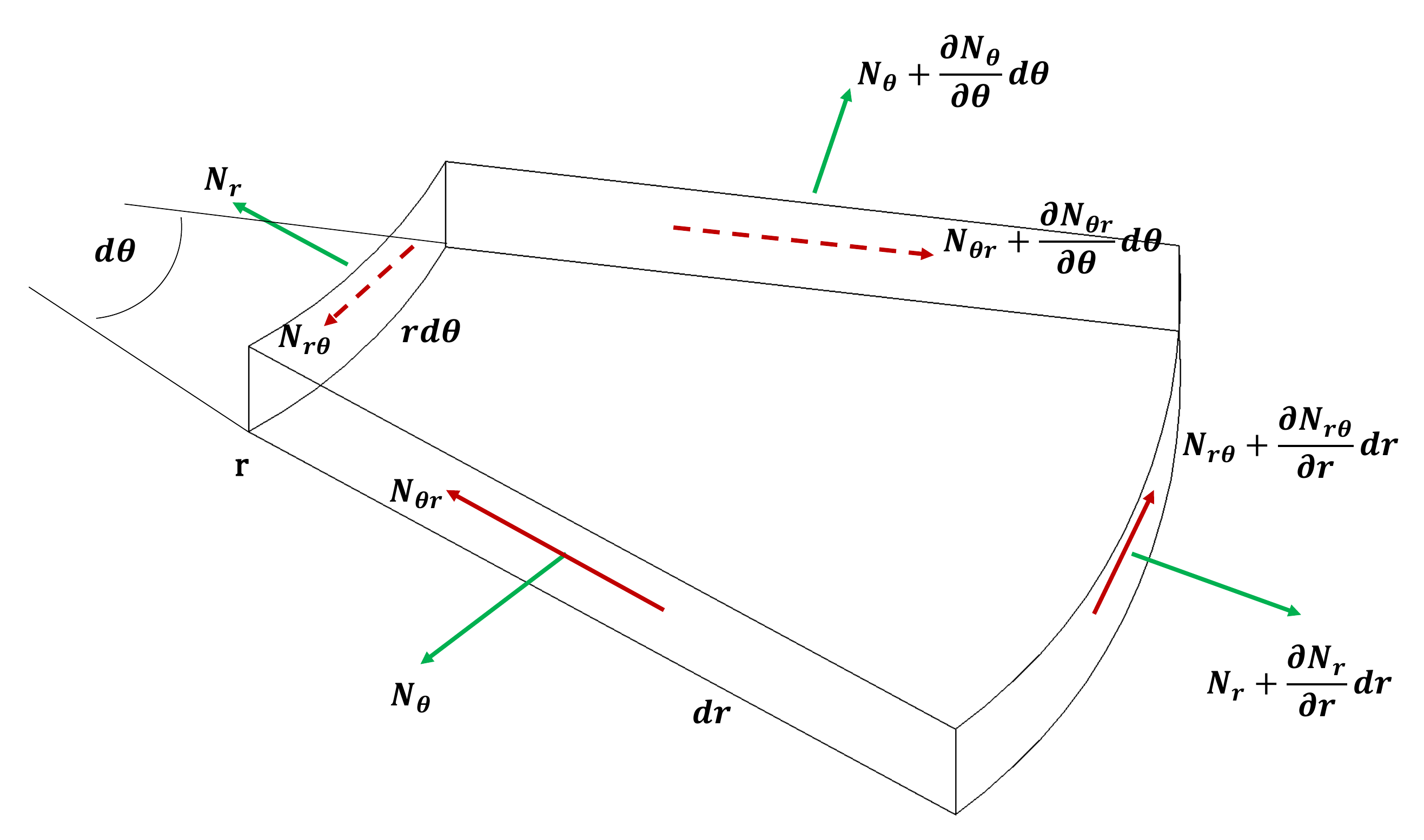}
    \caption{Infinitesimal plate element for analysis of in-plane plate vibration in polar coordinates for a circular plate undergoing axisymmetric radial vibrations.)}
    \label{fig:plate element}
\end{figure}

Considering force equilibrium condition along the radial direction $(r)$ in the infinitesimal plate element as shown in Figure~\ref{fig:plate element}, we obtain:
\begin{equation}
    \frac{\partial N_r}{\partial r}+\frac{1}{r}\frac{\partial N_{r\theta}}{r}+\frac{N_r-N_\theta}{r}=\rho h \frac{\partial^2 u_r}{\partial t^2}
\end{equation}

For axisymmetric case, the equilibrium equation can be written as:
\begin{equation}
    \frac{\partial N_r}{\partial r}+\frac{N_r-N_\theta}{r}=\rho h \frac{\partial^2 u_r}{\partial t^2}
\end{equation}

From the expressions of radial and circumferential forces obtained in equations~\eqref{eq:A8} and \eqref{eq:A9} respectively, we obtain the following expression for governing equation of axisymmetric radial in-plane vibration:
\begin{equation}
    \frac{E}{1-\nu^2}\left(\frac{\partial^2 u_r}{\partial r^2}+\frac{1}{r}\frac{\partial u_r}{\partial r}-\frac{u_r}{r^2}\right)=\rho \frac{\partial^2 u_r}{\partial t^2}
\end{equation}

Considering harmonic motion, we can express $u_r(r,t)=U_r(r)e^{i\omega t}$, and
re-write the governing equation of motion (upon dividing all terms in the equation by $e^{i\omega t}$):
\begin{equation}\label{eq:A16}
    \frac{E}{\rho\left(1-\nu^2\right)}\left(\frac{d^2U_r}{dr^2}+\frac{1}{r}\frac{dU_r}{dr}-\frac{U_r}{r^2}\right)+\omega^2 U_r=0
\end{equation}

The general solution for equation~\eqref{eq:A16} is of the form:
\begin{equation}
    U_r(r)=A_1 J_1\left(\beta \frac{r}{a}\right)+A_2 Y_1\left(\beta \frac{r}{a}\right)
\end{equation}
where,
\begin{equation}
    \beta=\sqrt{\frac{\rho a^2 \left(1-\nu^2\right)}{E}}\times \omega
\end{equation}
and $J_1(\frac{\beta r}{a})$ is Bessel function of first kind of order one and $Y_1(\frac{\beta r}{a})$ is Bessel function of second kind of order one.
The constants $A_1$ and $A_2$ are determined from appropriate boundary conditions.

\section{Effect of inertia of lateral motion on axisymmetric vibration of thick plates}
\label{app:b}

For modeling the effect of inertia of lateral motion on axisymmetric vibration, a thick rectangular plate undergoing axisymmetric extensional vibration is analyzed and the solution is then extended to circular plates.
The displacement field along $x$, $y$ and $z$ for a rectangular plate undergoing axisymmetric radial vibration with lateral displacement in the $z-$direction are expressed as:

\begin{gather} 
u_x(t)=u(x,t) \\ 
u_y(t)=0 \\
u_z(t)=-\nu z \frac{\partial u(x,t)}{\partial x}
\end{gather}

The kinetic energy of the plate can be expressed as:
\begin{equation} \label{eq:KE}
\begin{split}
T &=\frac{1}{2}\int_{0}^L{\rho A (\frac{\partial u}{\partial t})^2}dx + \frac{1}{2}\int_{0}^L{\rho \nu^2 \iint{z^2 dA \left(\frac{\partial^2 u}{\partial x \partial t}\right)^2}}dx \\
 & = \frac{1}{2}\int_{0}^L{\rho A (\frac{\partial u}{\partial t})^2}dx + \frac{1}{2}\int_{0}^L{\rho \nu^2 I_{zz} \left(\frac{\partial^2 u}{\partial x \partial t}\right)^2}dx
\end{split}
\end{equation}
where $\rho$ denotes the material density, $A$ denotes area of cross-section, and $I_{zz}$ is the area moment of inertia of the plate along the $z-$direction (thickness).
By extended Hamilton's principle, we obtain the governing equation for radial vibration including the effect of lateral displacement as:
\begin{equation}
    \rho \nu^2 I_{zz}\frac{\partial^4 u(x,t)}{\partial x^2 \partial t^2}+\frac{EA}{1-\nu^2}\frac{\partial^2 u}{\partial x^2}=\rho A \frac{\partial^2 u(x,t)}{\partial t^2}
\end{equation}

Assuming harmonic vibrations, with solution for displacement of form $u(x,t)=U(x)\cos(\omega t)$, we can re-write the governing equation as:
\begin{equation}
    \left(\frac{EA}{1-\nu^2}-\rho \omega^2 \nu^2 I_{zz}\right)U"(x)+\rho A \omega^2 U(x)=0
\end{equation}
Thus, we obtain the modified dimensionless frequency parameter as:
\begin{equation}
    \beta=\sqrt{\frac{\rho L^2}{\frac{E}{1-\nu^2}-\frac{\rho \\ \omega^2 \nu^2 I_{zz}}{A}}} \times \omega
\end{equation}

Extending this solution to circular thick plates, the modified dimensionless frequency parameter is expressed as:

\begin{equation}
    \beta=\sqrt{\frac{\rho a^2}{\frac{E}{1-\nu^2}-\frac{\rho \omega^2 \nu^2 I_{zz}}{A}}} \times \omega
\end{equation}

Re-arranging the above equation, we obtain the relation:
\begin{gather}
    \frac{E}{\rho a^2 (1-\nu^2)}=\left(\frac{1}{\beta^2}+\frac{\nu^2I_{zz}}{Aa^2}\right)\omega^2 \\
    \frac{E}{\rho a^2 (1-\nu^2)}=\frac{\omega^2}{\beta_m^2}
\end{gather}
where
\begin{equation}
    \frac{1}{\beta_m^2}=\frac{1}{\beta^2}+\frac{\nu^2 I_{zz}}{A a^2}
\end{equation}

For a circular disc, the area moment of inertia along the z-axis $(I_{zz})$ is not constant. Therefore an optimal solution was calculated by minimizing the error between analytically obtained and FEM simulation result for resonance frequency of a solid disc with increasing thickness to radius ratio. Using the optimized moment of inertia factor, the modified dimensionless frequency parameter incorporating lateral displacement is expressed as:
\begin{equation}
    \frac{1}{\beta_m^2}=\frac{1}{\beta^2}+\frac{\nu^2 \times 3.93568 \times h^2}{12 \times a^2}
\end{equation}
This relation is used for the calculation of natural frequencies for a thick circular disc with a square PZT transducer affixed on one of the circular faces.

\bibliography{biblo}

\end{document}